\documentclass[preprint,authoryear,12pt]{elsarticle}

\usepackage{epsfig}
\usepackage{color}
\usepackage{amssymb}

\journal{Advances in Space Research}

\newcommand{\HESS}{H.E.S.S.\/ }

\def\grays{$\gamma$~rays\ }
\def\gray{$\gamma$~ray\ }
\begin{document}

\begin{frontmatter}



\title{The Galactic Sky seen by H.E.S.S.}


\author{Mathieu de Naurois\corref{cor}}
\address{Laboratoire Leprince-Ringuet, Ecole Polytechnique, CNRS/IN2P3, F-91128 Palaiseau, France}
\cortext[cor]{Corresponding author}
\ead{denauroi@in2p3.fr}


\author{for the H.E.S.S. collaboration}


\begin{abstract}

The H.E.S.S. experiment is an array of four imaging Cherenkov telescopes
located in the Khomas Highlands of Namibia. It has been operating in its full configuration since
December 2003 and detects very-high-energy (VHE) \grays ranging from 
$100$~GeV to $\sim 50$~TeV.
Since 2004, the continuous observation of the Galactic Plane by the
H.E.S.S. array of telescopes has yielded the discovery of more than 50 
sources, belonging
to the classes of pulsar wind nebulae (PWN), supernova remnants (SNR),
\gray binaries and, more recently, a stellar cluster and molecular 
clouds
in the vicinity of shell-type SNRs. Galactic emission seen by H.E.S.S.
and its implications for particle acceleration in our Galaxy are
discussed.

\end{abstract}


\end{frontmatter}

\parindent=0.5 cm

\section{Introduction}

In the last decade, the third generation of imaging atmospheric Cherenkov telescope (H.E.S.S., VERITAS, MAGIC
and CANGAROO-III) came into operation and opened up a previously largely unexplored window on the very-high-energy (VHE) Universe. 
Since the pioneering Whipple experiment in 1989, the sensitivity has been increased by a factor of 100, leading to a 
detection time of $\sim 25~\mathrm{s}$ for a source of the intensity of the Crab Nebula compared to $50~\mathrm{h}$ for the 
original detection \citep{whipple-crab}. The TeV source catalgue now comprises more than $100$ sources\footnote{See TeVCat, an online TeV $\gamma$-ray catalog, at http://tevcat.uchicago.edu/}.
This scientific breakthrough was made possible by the combination of telescopes with large mirrors, cameras with fast photo-detectors and fine pixelation and 
the sterescopic observation technique which, by combining several views of the same shower seen from 
different telescopes, allows a simple geometric reconstruction of the direction of the primary gamma ray and a significant improvement 
of the angular resolution and rejection capabilites. 
The High Energy Stereoscopic System (H.E.S.S.), an array of four imaging atmospheric Cherenkov telescopes 
situated in the Khomas Highland of Namibia \citep{hess-crab}, played a major role in the opening up of this field, with in particular
a very successful systematic survey of the inner parts of the Galactic Plane starting in 2004 and  extended continously
since then \citep{aharonian2005b,hessgalaxy, chaves2009}. The angular resolution of \HESS is better than $0.1^\circ$ at all 
the accessible energies from $\sim 120~\mathrm{GeV}$ to several tens of TeV
and the energy resolution is about $15\%$ for a threshold varying from $\sim 120~\mathrm{GeV}$ at zenith to about $700~\mathrm{GeV}$
at a zenith angle of $60^\circ$. The sensitivity of the \HESS instrument was recently improved by a factor of $\sim 2$ by the developement 
of more sophisticated analysis techniques \citep{ohm2009,dubois2009,denaurois2009,naumann2009,fiasson2010}, 
some of which also yielded an angular resolution improved by about $30\%$.

The \HESS Galactic Plane survey now covers most of the Galactic Plane as seen from 
the Southern Hemisphere, and led to the discovery of a rich population of more than 50 Galactic sources, belonging to 
the classes of pulsar wind nebulae (PWN), supernova remnants (SNR), \gray binaries and, more recently, stellar clusters 
and molecular clouds in the vicinity of shell-type SNRs.

Outside of the Galactic Plane, more than 30 point-like sources have been discovered and associated with active galactic nuclei (AGN),
mostly objects of the BL Lacertae type (BL Lac).

\section{Survey of the Galactic Plane}

The \HESS Galactic Plane survey (GPS) has been a core component of the observation program since 2004. The original GPS \citep{aharonian2005b},
consisting of $\sim 230\mathrm{h}$ of observation after standard run-quality selection, covered the inner part of the Galaxy, from the Norma
to the Scutum-Crux spiral arms tangent ($l\pm 30^\circ$ in longitude and $b\pm 3^\circ$ in latitude). It resulted in the
firm discovery of eight previously unknown sources of VHE \grays with a statistical significance above $6 \sigma$ 
(post-trials\footnote{Since the GPS contains a large number of test positions, the significance has to be
corrected according to a ``trial factor.'' This trial factor accounts
for the increased probability of finding a fake signal with an increased
number of test positions for which a significance is calculated. A Monte Carlo simulation is
used to correct for the number of trials.})
and six likely sources above $4 \sigma$, all of them confirmed by subsequent deeper observations.

Between 2005 and 2009, the GPS was  extended significantly in longitude, from $l \sim -60^\circ$
to $l \sim  275^\circ$ \citep{chaves2009}. In addition, the overall exposure along the Galactic Plane was significantly increased
with more than 1400 hours of accumulated data (representing roughly one third of the total \HESS data set).
The \HESS exposure inside the Galactic Plane varies from a few hours on the less observed area to more than 100 hours in the 
deep exposure regions centered around targets of specific interest such as Sgr A*, RX~J1713.7-3946, or LS 5039, leading to a  sensitivity varying
between less than $1\%$ to about $10\%$ of the Crab Nebula flux.

\begin{figure}
\begin{center}
\includegraphics*[height=\textwidth,angle=-90]{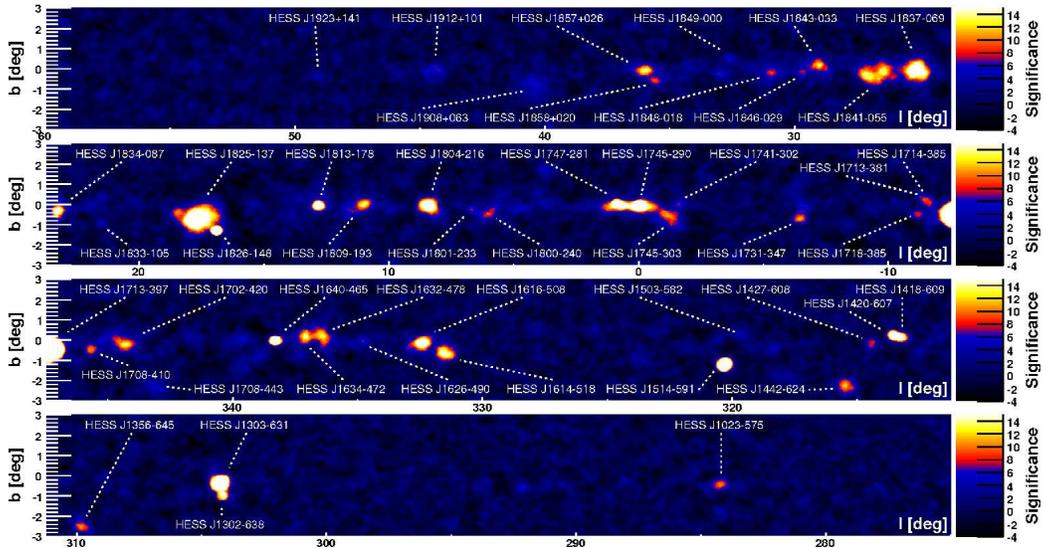}
\end{center}
\caption{\label{fig:scan}Pre-trial significance in the \HESS Galactic Plane Survey. Significance is truncated above 15 $\sigma$ to increase visibility. 
 Figure reproduced from \citet{chaves2009}.}
\end{figure}

The pre-trials significance map of the Galactic Plane, reproduced from \citet{chaves2009} and calculated
using the ring-background subtraction technique \citep{hess-crab} and {\it hard } cuts, is shown in Fig \ref{fig:scan}.
A total of 56 Galactic sources are detected in the GPS. The major population consists of PWN (29 identified 
sources) followed by SNR (9 associations) and binary systems (3 systems). 

Most of the Galactic VHE sources are found to be significantly extended, with sizes greater than the $\sim0.1^{\circ}$ \HESS point spread function (PSF). 
The few sources in the Galactic Plane that appear point-like are associated with young pulsar wind nebulae (PWN), including the Crab Nebula \citep{hess-crab},
or with VHE \gray  emitting high-mass X-ray binaries (HMXB) which include the very well established 
binaries PSR~B1259-63 \citep{psrb1259} and  LS~5039 \citep{hess-ls5039}. 
The point-like VHE source HESS~J0632+057 is now a strong candidate for a HMXB 
system following a recent multi-wavelength campaign \citep{discoveryhessj0632,hinton2009}.

After excluding five sources well off the Galactic Plane, with $\left|b\right| > 2^\circ$ ( HESS~J1356-645, HESS~J1442-624, 
HESS~J1507-662, HESS~J1514-591, and SN~1006), the latitude distribution of the Galactic sources is very 
narrow ($\langle b \rangle = -0.26^\circ$ with an RMS of $0.40^\circ$).
This scale is significantly smaller than the width of the region of significant \HESS exposure (of the order of $\sim 2^\circ$ in RMS),
and similar to the scale of the  molecular gas distribution. 
The latitude distribution is, at the first glance, compatible 
with what is the presumably parent populations of SNRs \citep{Green2004} and high spin-down luminosity pulsars
($\dot E > 10^{34}~\mathrm{ergs}~ \mathrm{s}^{-1}$) from \citet{atnf2005}.

\section{Supernova Remnants}

Expanding shock waves in SNRs are believed to be able to accelerate cosmic rays (CR) up to 
multi-TeV energies through the mechanism of diffusive shock acceleration (DSA) \citep[e.g.][]{DruryDSA}.
Moreover it was realized very early that, if a fraction of about $10\%$ of their explosion 
energy is converted into cosmic rays, SNRs are capable of maintaining the galatic 
CR flux at the observed level \citep{Baade1934}. Most shell-type SNRs are non-thermal radio emitters,  confirming
that electrons are accelerated up to at least GeV energies.
For a recent review on diffusive shock acceleration in the context of SNRs, see e.g. \citet{HillasCRReview}.

Four shell-like SNRs with clear shell-type morphology resolved
in VHE \grays have been detected by H.E.S.S.,  allowing direct investigation of the SNRs
as sources of cosmic rays. They are all remnants of recent  supernovae (less than a few kyr):
RX~J1713.7-3946 \citep{RXJ1713Nature,RXJ1713Deep}, RX~J0852.04622 - also known as Vela~Junior -  \citep{hessVelaJunior},
 SN~1006 \citep{hessSN1006} and HESS~J1731-347 \citep{hess1731}. 
A fifth case, RCW~86 \citep{hessRCW86}, might be added to this list although the TeV shell morphology has not 
yet been clearly proved.

All of them show a very clear correlation between
non-thermal X-ray emission and VHE \grays emission.


\subsection{Probing the acceleration mechanisms}

\begin{figure}[htb]
\begin{center}
\includegraphics*[width=0.75\textwidth,angle=0]{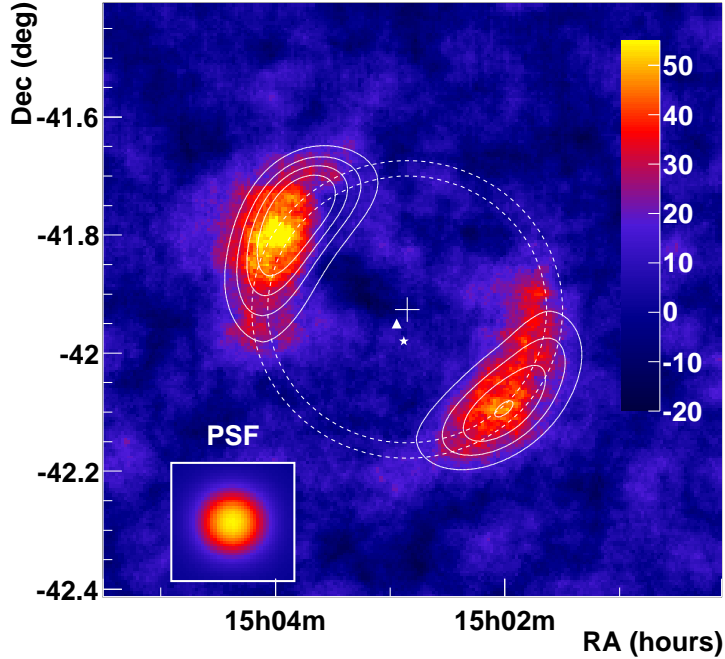}
\end{center}
\caption{\label{fig:sn1006map}HESS \grays correlated excess map of SN~1006. The linear colour scale is in units of excess counts per $\pi \times (0.05^\circ)^2$.
The white contours correspond to a constant X-ray intensity as derived from the XMM-Newton flux map and smoothed to the H.E.S.S. 
point spread function, enclosing respectively  80\% ,  60\% ,  40\%  and  20\% of the X-ray emission. Reproduced from \cite{hessSN1006}. }
\end{figure}

Amongst the aforementioned SNRs, SN~1006 is, due to its position 500 pc above the Galactic Plane,
an ideal case to study particle acceleration mechanisms. It indeed expands into a relatively uniform,
low density ($n\sim 0.085~\mathrm{cm}^{-3}$) medium \citep{acero2007,katsuda2009} and uniform magnetic field.
Moreover, SN~1006 is one of the best-observed SNRs with a rich data-set of radio, X-ray and optical measurements.
The progenitor of SN~1006 is believed to be a type Ia supernova \citep{Schaefer1996}, probably the brightest supernova
in recorded history. SN~1006 was detected by \HESS after a deep observation (130 h of data) and clearly exhibits 
a bi-polar morphology (Fig. \ref{fig:sn1006map}), strongly correlated with the non-thermal emission 
measured by XMM-Newton \citep{Rothenflug2004}. 

\begin{figure}[htb]
\begin{center}
\includegraphics*[width=0.85\textwidth,angle=0]{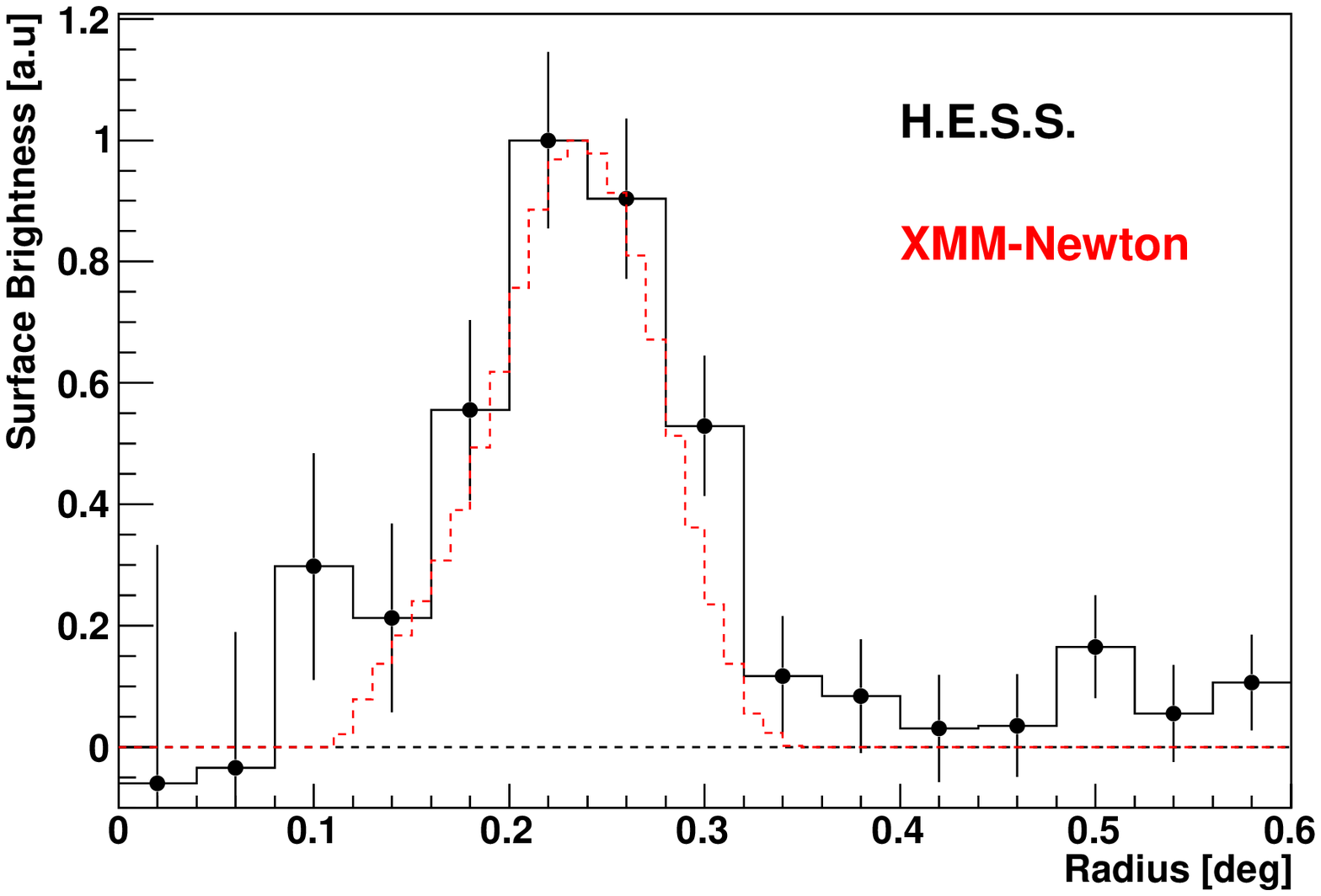}
\end{center}
\caption{\label{fig:sn1006profile}Radial profile around the centre of the SNR obtained from \HESS data and XMM-Newton data in the 2 - 4.5 keV energy band smoothed to \HESS. PSF.
Reproduced from \cite{hessSN1006}. }
\end{figure}

The close correlation between X-ray and VHE-emission demonstrated by the radial profile  (Fig. \ref{fig:sn1006profile}) 
points toward particle acceleration in the  strong shocks revealed by the Chandra observation of X-ray filaments \citep{Bamba2003}.
Moreover, the bipolar morphology of the VHE emission in the NE and SW regions of the remnant supports a major result of 
diffusive shock acceleration theory, according to which efficient downstream injection of suprathermal  charged nuclear ions 
is only possible for sufficiently small angles between the ambient magnetic field and shock normal.
Assuming a relatively uniform magnetic field oriented in the NE-SW direction, a higher 
density of accelerated nuclei at the poles is predicted \citep{Ellison1995,Malkov1995,Voelk2003}. 

\begin{figure}[htb]
\begin{center}
\includegraphics*[width=0.95\textwidth,angle=0]{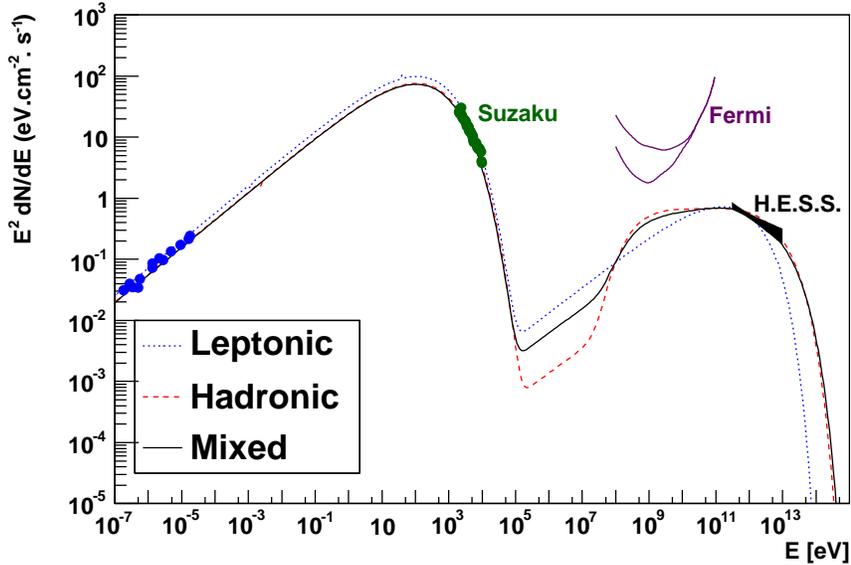}
\end{center}
\caption{\label{fig:sn1006sed}Spectral Energy Distribution predicted for SN~1006 for three different emission models: a purely
leptonic model (dotted blue line), an hadronic model (dashed red line) and a mixed model (solid black line). }
\end{figure}

Three different models were investigated to account for the spectral energy distribution (SED), and are compared in Fig. \ref{fig:sn1006sed}.
In a purely leptonic model (dotted blue line), TeV emission results from inverse-Compton scattering of multi-TeV electrons.  
The resulting magnetic field then needs to be higher than $30~\mathrm{\mu G}$ so that the IC emission does not exceed the measured VHE-flux.
In a second, dominantly hadronic model (dashed red line), TeV emission results from proton-proton interactions with $\pi^0$-production and subsequent decay,
whereas the X-ray emission is still produced by leptonic interactions. The resulting magnetic field  needs to be higher ($\sim 120~\mathrm{\mu G}$), 
which is consistent with magnetic field amplification at the shock, as indicated by the measurements of thin X-ray filaments mentioned above,
but requires that a very high fraction (about 20\%) of the supernova energy was converted into high-energy protons to account for
the TeV emission.
A mixed model (solid black line), in which hadronic and leptonic processes contribute almost equally to the very high-energy emission,
gives a good description of data, with a more reasonable overall acceleration efficiency of $\sim 12\%$.
Discrimination
between the different models might be possible  from observations in the MeV domain, however no instrument with 
the required sensitivity is currently in operation.

\subsection{Measurements of acceleration efficiency and magnetic fields}

The efficiency of particle acceleration in SNRs is an essential element to understand
whether SNRs can account for the observed level of CRs in the Galaxy. 
Elaborated, explicitly time-dependent, nonlinear kinetic models of cosmic ray (CR) acceleration in SNRs \citep[e.g.][]{Berezhko2009}
predict significant retroaction of the accelerated particles on the shock structure \citep{Berezhko1999} and result in magnetic field amplification.
Recent measurements using thermal Doppler broadening of the H$\alpha$ line on RCW~86 \citep{Helder2009} indicate that the postshock temperature 
is very significantly lower ($2.3\pm 0.3\mathrm{keV}$) than previously expected from the measured shock velocity ($42-70 \mathrm{keV}$). 
This was attributed to a  large acceleration efficiency, resulting in a cosmic-ray induced pressure that can
exceed the thermal pressure behind the shock.

In the last years, a growing consensus toward evidence for magnetic field amplification in the shocks of SNRs has emerged.
Measurements of thin filaments in SN~1006 with Chandra \citep{Berezhko2003,Bamba2003} strongly suggest rapid electron cooling in
intensified magnetific fields of the order of $0.1~\mathrm{mG}$, although they do not exclude alternate possible explanations such as field
damping.  More recently, the discovery of the brightening and decay of X-ray hot spots in the shell of the SNR RXJ1713.7-3946 on a one-year timescale
\citep{Uchiyama2007} indicates magnetic field amplification factors of the order of 100.

In summary, young supernova remnant have been proved to accelerate particles (electrons and/or protons) up to $100~\mathrm{TeV}$ at least. 
\grays emission at $100~\mathrm{TeV}$ is difficult to achieve with inverse-Compton scattering due to 
Klein Nishina suppression of the cross section at high energies.
Moreover, there is growing evidence that the efficiency can be as large as $50\%$ due to the retroaction of cosmic rays on the shock.
Magnetic field amplification, predicted in the framework of DSA, is now convincely observed
in several objects, thus further supporting the hypothesis of hadron acceleration in SNRs. The debate remains
unconclusive however, with comprehensive time-dependent models based on non-linear diffusive shock 
acceleration such as that of \citet{Morlino2009} arguing strongly in favor of hadronic acceleration, whereas other
authors \citep{Ellison2010}, using a consistent calculation of thermal X-ray emission, predict very intense X-ray thermal 
bremsstrahlung emission in the hadronic scenario, in contradiction with observations.

\subsection{Gamma-ray emission from illuminated clouds}

Further insight into the acceleration mechanisms in SNRs can come from the observation of middle-aged
supernova remnants in the vicinity of dense molecular clouds \citep{Gabici2009}. Indeed, production
of \grays is expected both during the acceleration phase of cosmic rays in the SNR shock and during
their subsequent propagation. Molecular clouds can act as cosmic-ray barometers, with an enhanced \gray emission 
proportional to the cloud mass, whereas inverse-Compton emission from accelerated electrons would not
be enhanced. Furthermore, the \gray emission from a cloud depends on the propagation time
and can therefore last much longer than the emission from the SNR, making the detection of clouds 
more probable \citep{Gabici2007}. A specific spectral signature, in the form of a concave energy spectrum,
has been predicted by \citet{Gabici2009} but not observed so far.

Four old SNRs in the vicinity of molecular clouds have been detected by H.E.S.S.: W~28 \citep{HESSW28}, an old ($\sim 35 - 150 \mathrm{kyr}$)
mixed-morphology SNR, HESS~J1745-303 \citep{HESSJ1745}, HESS~J1714-385 \citep{HESSCTB37A} and more recently 
HESS~J1923+141 \citep{hessw51}. In W~28, \HESS data show VHE emission from four different spots coincident with HII regions and 
dense molecular clouds revealed by NANTEN ${}^{12}\mathrm{CO}(J=1-0)$ data \citep{NANTENW28}. Interaction with a molecular
cloud  along its north and northeastern boundaries is further confirmed by the high concentration of
1720~MHz OH masers \citep{ClaussenMasersW28}. Under the assumption
of cloud distance between 2 and 4~kpc, a cosmic-ray overdensity in the range 13 to 32 is derived, which is consistent with
expectations. 

\begin{figure}
\begin{center}
\includegraphics*[width=0.8\textwidth,angle=0]{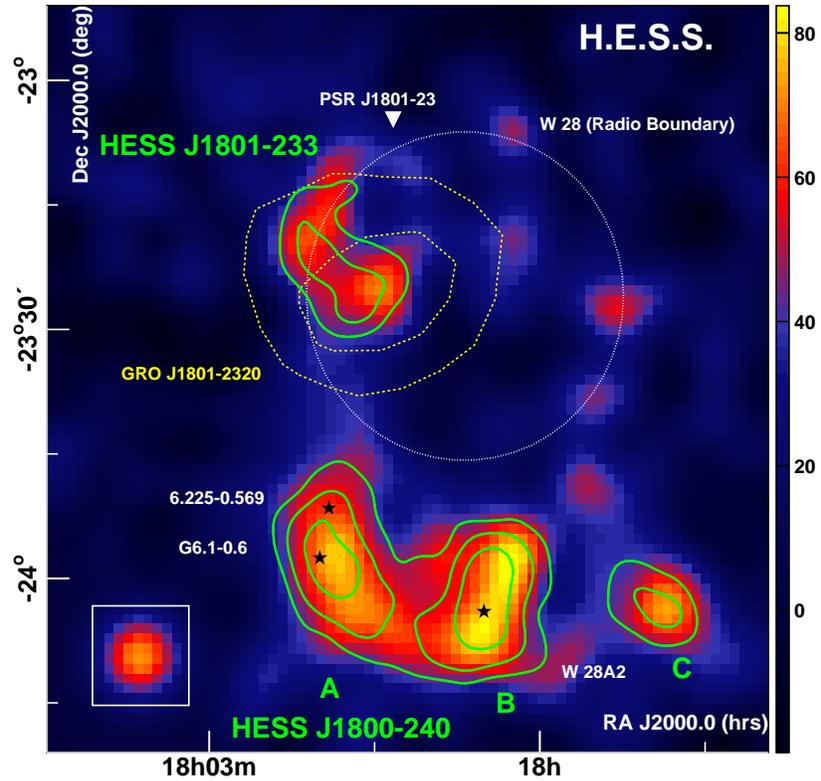}
\end{center}
\caption{\label{fig:w28map}Image (1.5$^\circ\times 1.5^\circ$) of the VHE $\gamma$-ray excess counts (events), corrected for 
    exposure and smoothed with a Gaussian of radius 4.2$^\prime$ (standard deviation). 
    Overlaid are solid green contours of VHE excess 
    (pre-trial) 
    significance levels of 4, 5, and 6$\sigma$, after integrating events within an oversampling radius $\theta$=0.1$^\circ$ 
    appropriate for pointlike sources. From  \citep{HESSW28}.}
\end{figure}

Similar conclusions are derived from observations of HESS~J1745-303, HESS~J1714-385 and HESS~J1923+141, thus
confirming that shell-type SNRs are efficient accelerators of hadronic cosmic rays. Three of these old SNRs
in the vicinity of clouds are firmly associated with bright Fermi sources \citep{FERMICatalog}:
W~28 is associated with 1FGL~J1801.3-2322c and 1FGL~J1800.5-2359c, HESS~J1714-385 is associated with 1FGL~J1714.5-3830
and HESS~J1923+141 with 0FGL~J1923.0+1411. A possible, less convincing, association also exists for HESS~J1804-216.
Spectral continuity between GeV and TeV energies, large cosmic-ray 
densities derived from the VHE flux and coincidence with the shocked clouds  strongly
favours a hadronic origin of the GeV and TeV $\gamma$~rays.

Further multiwavelength studies with ACTs and Fermi will help to understand
the mechanisms at the origin of Galactic CRs.

\section{Pulsar Wind Nebulae}

Nearly half of the Galactic VHE sources, starting from the famous Crab Nebula \citep{whipple-crab} 
are associated with young, energetic pulsars. These sources exhibit strong ultra-relativistic winds
of particles that lead to the formation of a synchrotron nebula when the winds interact with the surrounding
medium. Strong shocks are formed, resulting in acceleration of particles up to hundreds of TeV.

\HESS has detected a wide range of PWNs. Young PWNs such as the Crab Nebula \citep{hess-crab}, 
SNR~G~0.9+0.1 \citep{hessg09}, SNR~G~21.5-0.9 \citep{hess-youngpwn}, Kes~ 75 \citep{hesskes75}, 
 MSH~15-52 \citep{hessmsh1552} and  HESS~J1813-178 \citep{hessgalaxy,hess1813} are 
generally compact and unresolved. In such systems, 
VHE emission is generally attributed to inverse-Compton scattering of $1-100~\mathrm{TeV}$ electrons
\citep[e.g.][]{Atoyan1996}. Statistical studies \citep{2008AIPC.1085..698W} indicate that the majority, if not all, of young
high-spindown luminosity pulsars produce significant TeV emission.

\begin{figure}
\begin{center}
\includegraphics*[width=0.7\textwidth,angle=0]{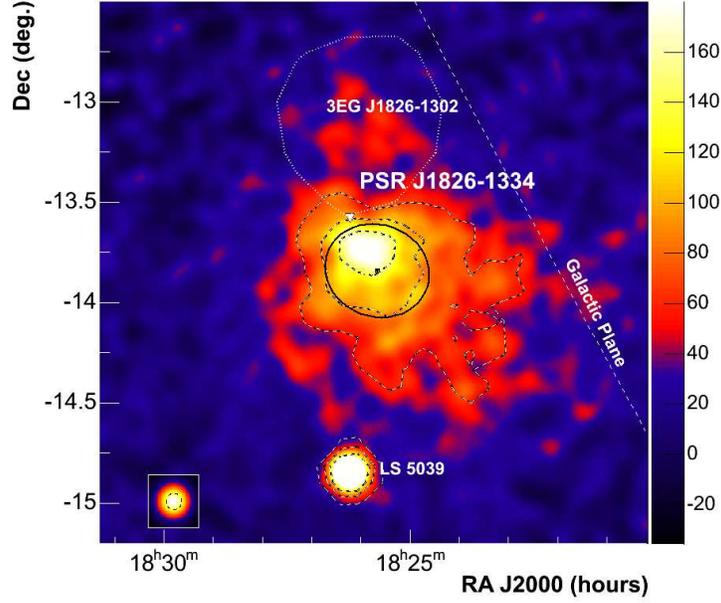}
\end{center}
\caption{\label{fig:hessj1825map}Acceptance-corrected smoothed excess map (smoothing radius
  $2.5'$) of the $2.7^\circ\ \times 2.7^\circ$ field of view
  surrounding HESS~J1825-137. From \citet{hess1825}.}
\end{figure}

\begin{figure}
\begin{center}
\includegraphics*[width=0.95\textwidth,angle=0]{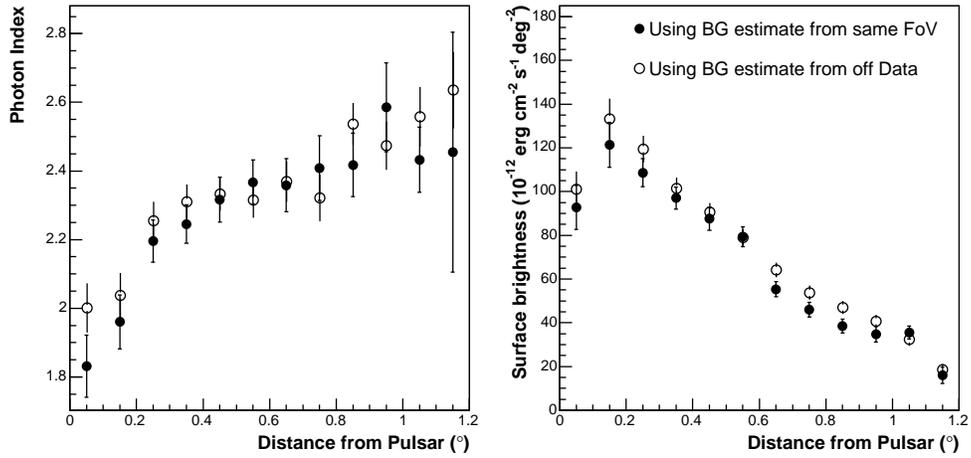}
\end{center}
\caption{\label{fig:hessj1825cooloing}Energy spectra of HESS\,J1825--137 in radial bins.  {\bf Left:} Power-law
  photon index as a function of the radius of the region (with respect
  to the pulsar position). {\bf Right:} Surface brightness between 0.25 and 10 TeV per
  integration region area in units of $10^{-12} \mathrm{erg} \,
  \mathrm{cm}^{-2} \, \mathrm{s}^{-1} \, \mathrm{deg}^{-2}$ as a
  function of the distance to the pulsar position. }
\end{figure}

Older PWNs such as HESS~J1825-137 \citep{hess1825} and HESS~J1303-631 \citep{hess1303pwn}
show more complex morphologies, with significant offsets between the pulsar and the
nebula. Spectral steepening away from the pulsar, detected in HESS~J1825-137 (Fig. \ref{fig:hessj1825map}
and \ref{fig:hessj1825cooloing}, is the first direct evidence of radiative cooling of electrons. 
VHE-emitting electrons are usually less energetic than those emitting X-rays, do not
suffer from severe radiative losses and therefore can survive in greater number from the early
epoches of the PWN evolution.  The high VHE luminosity of HESS~J1825-137 compared to the X-ray
luminosity can be explained by a significant contribution of `relic' electrons released in the early history 
of the pulsar, when the spin-down luminosity was higher.  
The variation of index with distance from the pulsar is attributed both to inverse-Compton and synchrotron
cooling of the continuously accelerated electrons.
Further confirmation of this mechanism is required from observation of other old PWNs and may
arise from HESS~J1303-631. Recently, the first extragalactic PWN
was detected in the Large Magellanic Cloud \citep{hesslmc}.

\section{Dark Accelerators}

About half of the
Galactic VHE sources show no clear counterparts in lower-energy wavebands and remain unidentified \citep{hessunid,TibollaFermi}, 
a  fraction similar to that observed with EGRET \citep{EGRETCatalog} and now with Fermi \citep{FERMICatalog}.
The vast majority of these sources are significantly extended, well beyond the PSF of H.E.S.S..
Understanding the emission mechanism powering these sources is a challenge of multi-wavelength astronomy.

It has been recently suggested  \citep{jagerpwn} that a significant fraction of these sources could
be old pulsar wind nebul\ae\  in a late evolution stage. Indeed, magneto-hydrodynamic simulations indicate that the magnetic field
in PWN decreases with time as $t^{-1/3}$, leading to the extinction of the synchrotron emission. In contrast, the
inverse-Compton emission tends to increase with time until most of the pulsar spindown power has been dumped into
the nebula.

Alternate models include, in particular, emission from molecular clouds illuminated by cosmic rays coming 
from a nearby source \citep{AharonianAtoyan1996,Gabici2009}, in which a concave energy spectrum
would constitute a definitive proof.

\section{Binary Systems}

X-ray binaries (XRBs) comprise a compact object such as a neutron star or black hole orbiting around a companion star. 
They are one of several types of astrophysical systems that  provide an environment in which the acceleration of 
particles and subsequent production of radiation might be periodic.
Modulation of this radiation, linked to the orbital motion of the binary system, provides key insights into 
the nature and location of particle acceleration and emission processes.

\begin{figure}[ht]
\begin{center}
\includegraphics*[width=0.8\textwidth,angle=0]{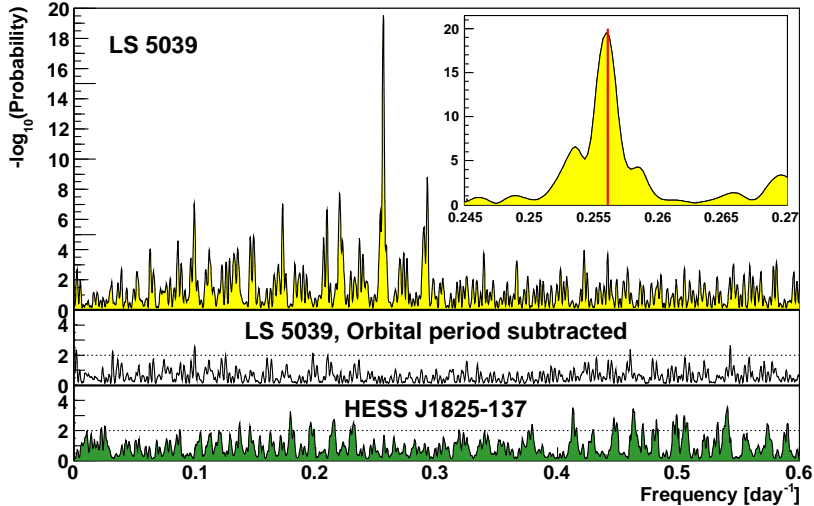}
\end{center}
\caption{\label{fig:ls5039period}{\bf Top:} Lomb-Scargle (LS) periodogram of the VHE runwise flux for LS~5039 (chance probability to obtain the LS power vs. frequency). 
      Inset: zoom around the highest peak (pre-trial probability $\sim$10$^{-20}$), which corresponds to a period of 3.9078$\pm$0.0015~days.
      {\bf Middle:} LS periodogram of the same data after subtraction of
      a pure sinusoidal component at the orbital period of 3.90603 days. 
      {\bf Bottom:} LS periodogram of the HESS source HESS~J1825$-$137 observed simultaneously in the same field of view. From \citet{hess-ls5039}}
\end{figure}

LS~5039 is one of the handful of X-ray binaries that have been
detected at VHE \grays \citep{hess-ls5039}.
Results from 70 hr of observations distributed over many orbital cycles
yielded a modulation  of the VHE \gray flux ($>100~\mathrm{GeV}$)  
with a period of $3.9078 \pm 0.0015~\mathrm{days}$ \citep{hess-ls5039paper2},
consistent with the orbital period reported by \citet{casares2005}.
The corresponding \HESS Lomb-Scargle periodogram is show in Fig. \ref{fig:ls5039period}.

\begin{figure}[htb]
\begin{center}
\includegraphics*[width=0.8\textwidth,angle=0]{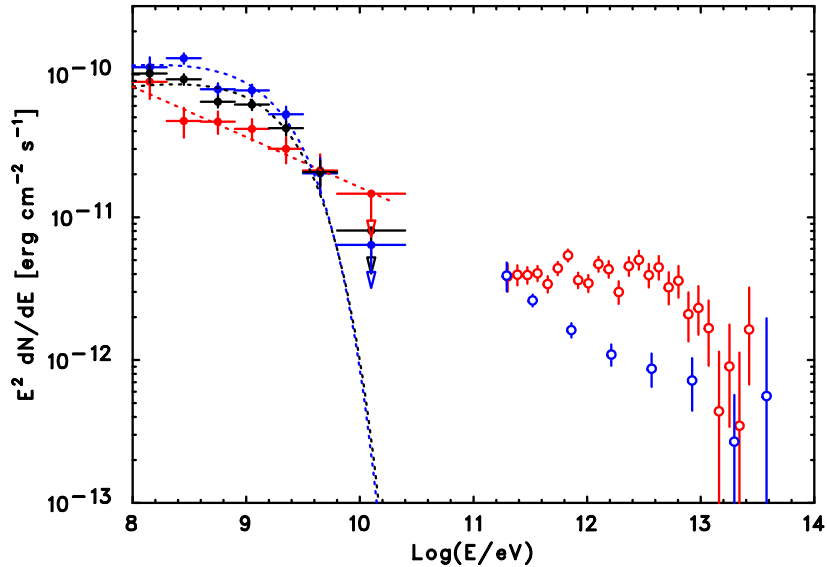}
\end{center}
\caption{\label{fig:ls5039spectrum}Phase resolved SED of LS~5039 with Fermi and \HESS. The black points (dotted line) represent the
phase-averaged Fermi/LAT spectrum. The red data points (dotted line) represent
the spectrum (overall fit) at inferior conjunction (Phase $0.45~0.$9); blue
data points (dotted line) represent the spectrum (overall fit) at superior conjunction
(Phases, $<0.45$ and $>0.9$). From \citet{fermi-ls5039}}
\end{figure}

Orbital modulation of the VHE spectrum, shown in Fig. \ref{fig:ls5039spectrum}, was interpreted
as the result of phase-dependent pair creation on the stellar photon field \citep[e.g.][]{maraschi1981,dubus2006}, 
leading to a significant absorption of the VHE flux at the superior conjunction (in blue on Fig. \ref{fig:ls5039spectrum}),
when the compact object is behind the star. 

Recent observations with Fermi \citep{fermi-ls5039} yielding a consistent period of $3.903 \pm 0.005 ~\mathrm{days}$
in GeV \grays, with the maximum of the GeV emission during the superior conjunction, when the TeV flux is at its minimum
and vice-versa (Fig. \ref{fig:ls5039spectrum}). This clear anticorrelation, predicted in inverse-Compton scattering
models  \citep[e.g.][]{dubuscerutti2008,cerutti2008}, results from the competition of the inverse-Compton
and pair-creation processes:
The inverse-Compton  emission in the MeV-GeV regime is enhanced when the highly-relativistic electrons seen by the observer encounter the seed
photons head-on (at superior conjunction) while VHE absorption due to pair production is maximum at the 
same position leading to suppression of VHE photons.

After the discovery of the binary pulsar PSR~B1259-63 by \HESS \citep{psrb1259,psrb1259b} and 
LS~I~+61~303 by MAGIC \citep{lsi61303}, a \HESS unidentified source, HESS~J0632+057 has
been recently added to the list of \gray binary candidates. \citep{discoveryhessj0632,hinton2009}.

\section{Massive Star Clusters}

Since they host supernova remnants and pulsar wind nebul\ae, massive stellar clusters
are obviously potential acceleration sites of VHE particles,  but several
alternate scenarios also provide plausible explanations for the production of VHE particles.

Many stellar clusters harbor massive stars frequently bound in multiple star systems.
The strong and fast winds form a collision region were particles can be accelerated up to \gray 
energies \citep[e.g.][]{1993ApJ...402..271E}. Acceleration models involve either leptonic processes 
(inverse-Compton scattering)
or hadronic processes (inelastic scattering of nucleons in the dense stellar wind followed
by production of neutral pions which subsequently decay into VHE $\gamma$ rays).

In addition, the strong stellar winds of individual massive stars also interact with each other and lead to the formation
of wind-blown bubbles, filled with a low-density hot plasma \citep{1982ApJ...253..188V} in which diffusive shock acceleration
can take place. \citep{2000APh....13..161K,2001SSRv...99..317B,2004AA...424..747P,2005ApJ...628..738H,2008PhST..132a4024D}. 

VHE \gray excess emission towards (at least) two massive star clusters has been detected: Westerlund~1 \citep{hess-westerlund1}, the most massive cluster in our Galaxy, and Westerlund~2 \citep{hess-westerlund2,hess-westerlund2-second}. 
In addition, two unidentified \HESS sources, HESS~J1614-581 and HESS~J1848-018, may be related to
similar systems \citep{2010ASPC..422..265O}.

Westerlund 1 harbors at least 24 Wolf-Rayet stars 
of which $>70\%$ are in binary systems \citep{2006AA...457..591G}, in addition to  blue and red super-giants stars.
This makes the collective winds scenario quite attractive.

In the Westerlund 2 region, the prominent binary system WR20a, including its colliding
wind zone \citep[e.g.][]{2005MNRAS.363L..46B}, would appear as a point source for observations with the \HESS telescope array.
After detection of extended  VHE \gray emission and apparent lack of flux variability over orbital timescales,
alternative mechanisms involving collective stellar wind effects are preferred. 
Furthermore, the recent discovery  of two neighboring bright pulsars 1FGL~J1023.0~5746 and 1FGL~J1028.4~5810 by the Fermi collaboration \citep{2009ApJ...695L..72A}
motivated again the consideration of the pulsar wind nebula emission.

The connection with collective stellar wind emission hypothesis appears quite attractive in these systems, in particular due to
the extension of the VHE emission, but needs to be confirmed through further observations in the \HESS energy band
and by considering the full MWL perspective. Unique signatures, such as energy dependent morphology changes or MWL correlation (or anticorrelation)
with the VHE emission are still needed to unequivocaly confirm the connection to wind-related phenomena in massive star clusters.

\section{Conclusions}

The results presented here represent only a subset of the discoveries
made by the ground based \gray astronomy, and in particular by \HESS during the
last years.  Other significant results from \HESS, regarding in particular extragalactic
sources of VHE $\gamma$ rays, include amongst others observation
of fast variability and giant flares in active galactic nuclei, detection of starburst 
galaxies or search of dark matter annihilation in gravitational potential 
wells.

In 2012, the H.E.S.S. array will be completed with the addition of a very large telescope
(28 m diameter) in the centre of the present array. An energy threshold of
30 GeV is expected from this large telescope in stand-alone mode. Stereoscopy 
including the very large telescope and at least one of the other four
should allow for a threshold of 80 GeV with improved sensitivity. 
This new facility will provide a significant 
overlap between the energy ranges covered by \gray satellites (Fermi-LAT
and AGILE) and H.E.S.S., thus helping to understand the mechanisms
at the origin of Galactic CRs.

\section{Acknowledgements}
The support of the Namibian authorities and of the University of Namibia
in facilitating the construction and operation of H.E.S.S. is gratefully
acknowledged, as is the support by the German Ministry for Education and
Research (BMBF), the Max Planck Society, the French Ministry for Research,
the CNRS-IN2P3 and the Astroparticle Interdisciplinary Programme of the
CNRS, the U.K. Science and Technology Facilities Council (STFC),
the IPNP of the Charles University, the Polish Ministry of Science and 
Higher Education, the South African Department of
Science and Technology and National Research Foundation, and by the
University of Namibia. We appreciate the excellent work of the technical
support staff in Berlin, Durham, Hamburg, Heidelberg, Palaiseau, Paris,
Saclay, and in Namibia in the construction and operation of the
equipment.

\bibliographystyle{custom} 
\bibliography{cospar2010_denauroi} 

\end{document}